\documentclass[aps,prl,twocolumn,reprint]{revtex4-1}

\usepackage{amsmath,amsfonts,color,graphicx,fullpage, subfig,setspace, floatrow,dcolumn}
\newfloatcommand{capbtabbox}{table}[][\FBwidth]
\usepackage[colorlinks={true},dvipdfm]{hyperref}
\usepackage[singlelinecheck=off,justification=raggedright]{caption}
\hypersetup{citecolor={blue}, filecolor={blue}, linkcolor={blue}, urlcolor={blue}}
\begin{document}
\date{\today}

\author{David Hocker}
\affiliation{Department of Chemistry, Princeton University, Princeton, NJ 08544, USA}
\author{Herschel Rabitz}
\affiliation{Department of Chemistry, Princeton University, Princeton, NJ 08544, USA}


\title{Invariance in multi-objective quantum control}
\begin{abstract}
Simultaneous optimization of multiple quantum objectives is often considered a demanding task. However, a special circumstance arises when  a primary objective is pitted against a set of secondary objectives, which we show leads to invariant behavior of the secondary objectives upon the primary one approaching its optimal value. Still, practical relationships among the objectives will generally lead to a threshold, beyond which system re-engineering is required to further increase the primary objective. This finding is of broad significance for reaching high performance in quantum technologies.
\end{abstract}
\date{\today}

\maketitle
The control of complex quantum phenomena \cite{brif_jnewphys_2010} can entail simultaneous optimization of multiple objectives in many applications. Nevertheless, a common circumstance is to seek maximization of some \emph{primary} measure $J$ to high fidelity. For example, in the emerging field of quantum technologies \cite{quantum_tech_review} $J$ may be defined as the success of a quantum information processing gate \cite{Nielsen_Chuang_qc}, or a measurement from a metrology or a quantum sensing experiment \cite{metrology_review, Kotler_nat_2014}. Additionally, while seeking high fidelity for $J$, there will typically be additional secondary objectives $\{K_i\}$ such as performing control with the least amount of energy, preserving coherence, or avoiding unwanted noise effects. Optimization of the primary and secondary objectives are often performed by seeking tailored, applied fields, and consideration of $J$ and   $\{K_i\}$ objectives will generally involve a tradeoff. However, this letter will point out an attractive feature of unitary quantum control imparting a special general character to this tradeoff in the desired high fidelity regime for $J$.

We assume that the physical circumstances demand that $J$ be nearly optimal, and we aim to discern the tradeoffs when additionally optimizing $\{K_i\}$. Characterizing such tradeoffs is commonly called the Pareto problem, where typically increase of one objective value through changing the controls cannot be performed without penalty to the other objectives \cite{deb_2001, coello_2007}. Such circumstances pose a difficult situation, as no operating condition may be satisfactory for meeting  all concerns  \cite{JPhysA.44.2011.095302, moore_pra_2012,PhysRevA.78.033414,  PhysRevA.85.032305, 1367-2630-11-1-013019}.

Here we demonstrate that rather than finding a tradeoff, at  \emph{high} values of $J$ an invariance settles in for the value of all secondary objectives $\{K_i\}$. This behavior holds for any well posed primary quantum objective $J$, and is important in the growing efforts to develop quantum technologies. To illustrate, we will demonstrate its implications in three physical circumstances (see eqs. \ref{eq:pif}-\ref{eq:J}) where high fidelity of some operation $J$ is necessary. Spin systems will be used in this regard. As the two objectives we take the fidelity of $J$  and a measure of robustness $K_{\beta}$ to noise disturbances of type $\beta$ in the system. Numerical calculations will show that for increasingly high fidelity of $J$, the robustness $K_{\beta}$ to either detuning noise or control noise freezes in at an invariant value. The same conclusions would apply to analogous optimization operation directly performed in the laboratory. In general, the invariance of $K_{\beta}$ to increasing fidelity of $J$ will reach a threshold, whereby proceeding further with $J$ will call for improving $K_{\beta}$ that can only be achieved by appropriate re-engineering of the system. Depending on the circumstances, dealing with infringing $\{K_i\}$ could call for re-engineering of the control sources, the physical system, the detectors, etc. While this invariant behavior is independent of search technique, it does provide algorithmic advantages, as even a modest ensemble of Monte-Carlo (MC) samples of nearly optimal controls can readily identify the best invariant value of $K_\beta$.


The invariant behavior is shown to arise due to the generally expected, unencumbered convergence toward the optimal value of $J$ along a path on the quantum control fidelity \emph{landscape}, defined as the functional mapping from the controls $\epsilon(t)$ to the objective $J$. As background, the success of a growing number of optimal control experiments led to the formulation of a key theorem, referred to here as the ``landscape principle," demonstrating that the topology of well-conditioned quantum control landscapes allows for facile determination of optimal controls \cite{Hsieh_pra_2008, Hsieh_jcp_2009, PhysRevA.74.012721}. This principle provides a foundation to explain the growing successes of quantum control \cite{quantphyslett_1_1_2012}, and at this juncture very few exceptions are known to this principle, thereby providing a solid basis for its practical exploitation \cite{PhysRevA.89.043421}.

The convergence of an optimal control $\epsilon(t)$ to produce a high value of $J$ is observed by parametrizing the control by a search variable $s \geq 0$. In this fashion, traversal over the landscape will start from $s=0$, where $J$ is generally at low fidelity, out  to a final value $s_f = \infty$ (\emph{i.e.}, in practice some finite value of $s_f$ where $J$ is near optimal, taken as $J=1$). A gradient-based optimization along $s$ will always seek to increase $J$ through
\begin{align}
\label{eq:dmorph1}
\frac{d J}{d s} &= \int_0^T \frac{\delta J}{\delta \epsilon(s,t)}\frac{\partial \epsilon(s,t)}{\partial s} dt \quad \geq 0,
\end{align}
which is guaranteed by \cite{Moore_pra_2011,Rothman_pra_2006}
\begin{align}
\label{eq:dmorph2}
\frac{\partial \epsilon(s,t)}{\partial s} &= \frac{\delta J}{\delta \epsilon(s,t)}.
\end{align}
Numerous simulations and experiments have demonstrated that the landscape principle enforces that norm of the gradient $\delta J/ \delta \epsilon(s,t)$ has a Gaussian-like shape as a function of $s$ that eventually decreases as $s \rightarrow s_f$ until a critical point at the landscape top is approached where $\delta J/ \delta \epsilon(s_f,t) \simeq 0$,  $\forall t$. The condition in eq. (\ref{eq:dmorph2}) implies that near optimal fidelity,
\begin{align}
\label{eq:sfinal}
\lim_{s \rightarrow s_f} \frac{\partial \epsilon(s,t)}{\partial s} \simeq 0,
\end{align}
where ever decreasing variations in $\epsilon(s,t)$ occur for $s \rightarrow s_f$. Consequently, ancillary objectives $K_i[\epsilon(s,t)]$ that solely depend upon the controls also converge at high $s$ values. 


\begin{figure}[h]
\subfloat{\includegraphics[width=0.9\textwidth, height=0.9\textheight, keepaspectratio]{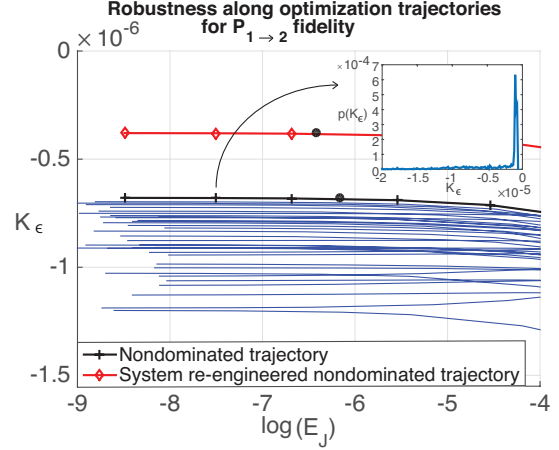}}
\caption{\label{fig:mcfront}  Color online. Field noise robustness measure in eq. (\ref{eq:K_a}) for representative collection of Monte Carlo optimization trajectories. Each curve arises from a single run up the $P_{1\rightarrow 2}$ landscape with a randomly chosen initial field. Convergence to high $J$ (\emph{i.e.} at smaller error $E_J$) demonstrates an invariance, evident in each line culminating in a single, non-dominated and increasingly flat trajectory for $K_{\epsilon}$ (labeled with + signs). Threshold point where $E_J=-K_{\epsilon}$ is given as a dot. The improvement of the threshold point is shown here by re-engineering noise statistics to increase $K_{\epsilon}$ (labeled with $\diamond$ signs).   Inset, the probability distribution $\text{p}(K_{\epsilon})$ of 2000 MC samples at $\log(E_J)=-7.5$, showing a preference to sample the largest invariant $K_{\epsilon}$ value. }
\end{figure}

While traditional Pareto ``front" tradeoffs will likely exist when \emph{initially} climbing the control landscape, the asymptotic behavior of \emph{all} individual search trajectories via eq. (\ref{eq:sfinal}) (\emph{i.e.}, each guaranteed to reach high fidelity for $J$ by the landscape principle) likewise assures a high degree of invariance between the fidelity of $J$ and secondary objectives. Figure \ref{fig:mcfront} demonstrates this global behavior by showing how a MC sampling of control trajectories climbing the landscape can readily reveal convergence to the highest \emph{invariant} value of field noise robustness $K_{\epsilon}$ at ever lower fidelity \emph{error} $E_J$ (\emph{i.e.}, higher $J$) given as $E_J=1-J$. Maximization of the secondary objective $K_{\epsilon}$ is seen as a single, flat curve labeled $+$ in the regime of high fidelity (\emph{i.e.}, $\log(E_J) \lesssim -6$) where $K_{\epsilon}$ plateaus \footnote{As this invariance is a consequence of the landscape, and not a particular search algorithm, any suitable search technique should identify invariant behavior}.

However, this invariance hides a competition, as a threshold value of $J$ will occur where $E_J = -K_{\beta}$ (Fig. \ref{fig:mcfront}, black dots for the highest value of $K_{\beta}$) , such that pushing further on fidelity can only be achieved by some suitable re-engineering of the system to increase $K_{\beta}$. Such an effect is demonstrated by the improved trajectory in Fig. \ref{fig:mcfront} ($\diamond$ signs), in which the noise statistics were re-engineered (\emph{i.e.}, a longer noise correlation time), leading to an increase in $K_{\beta}$ and an accompanying lowering of the threshold. Similar threshold behavior on \emph{any} secondary objective likely occurs in all quantum control problems (see Supplementary Material).


We consider three types of primary objectives:
\begin{align}
\label{eq:pif}
P_{i\rightarrow f} &= | \langle f | U(T)|i\rangle |^2 & \text{state} \\
\label{eq:tro}
\langle O \rangle &= \text{Tr}[\rho(T) \hat{O} ] & \text{observable}\\
\label{eq:J}
F_W &= 1-\frac{1}{4N}\| U(T)-W \|^2 & \text{unitary}
\end{align}
$J$ is taken as the objective in either eqs. (\ref{eq:pif}), (\ref{eq:tro}) or (\ref{eq:J}), which range here from worst fidelity  ($J=0$) to optimal (all normalized to $J=1$). Control of any $N$-level  closed quantum system manipulates the dynamics of a unitary propagator $U(t) \equiv U(t,0)$, and correspondingly either its state $|\psi(t)\rangle = U(t)|i\rangle$ or density matrix $\rho(t) = U(t) \rho_i U^{\dag}(t)$.  $U(t)$ is driven by a Hamiltonian $H(t)$ through solution of the Schr\"odinger equation with short time steps $\Delta t$,
\begin{align}
\label{eq:prop}
U(t+\Delta t) &= e^{ -\frac{i}{\hbar} H(t+\Delta t)  \Delta t}  U(t).
\end{align}
For our numerical demonstrations, $M$ coupled spins form the Hamiltonian. Each spin has a field-free drift term with transition energy $\omega_i$ between the spin states $|0\rangle = [1,0]^{T}$ and $|1\rangle = [0,1]^T$, a control field $\epsilon_i(t)$, and collectively isotropic Heisenberg coupling terms of strength $J_{i,j}$ linking the spins (dynamical details provided in Supplementary Material).



Maximization of the fidelity for two and four-level systems (\emph{i.e.} $M = 1$ or 2, respectively) are considered for the following objectives: $P_{1\rightarrow 2}$, $P_{1 \rightarrow 4}$, $\langle \sigma_x \rangle$ ($\rho_i = |0\rangle \langle 0|$), and $\langle \sigma_x^{(1)} \rangle$ ( $\rho_i = |00\rangle \langle 00|$ ), where $\langle \sigma_x \rangle$ and $\langle \sigma_x^{(1)} \rangle$ are all rescaled to range from 0 to 1, with 1 being optimal . For unitary transformation control in eq. (\ref{eq:J}) , we consider one-qubit Hadamard ($F_{\mathbf{H}}$) and two-qubit CNOT gates ($F_{\mathbf{CNOT}}$). Two level dynamics are simulated using eq. (\ref{eq:prop}) over the interval $T=1$ ($\Delta t=0.01$), while four level dynamics are performed over $T=6$ ($\Delta t=0.06$).  $\omega_1=20$, $\omega_2 = 24$, and a weak spin coupling is chosen as $J_{1,2} = 0.2$. Optimization is performed via the D-MORPH gradient algorithm by integrating eq. (\ref{eq:dmorph2}) with a fourth-order Runge Kutta integrator  (Matlab's $\mathbf{ode45}$) over a suitably long $s$ interval \cite{Moore_pra_2011,Rothman_pra_2006}.


A measure for robustness to noise that is valid at high fidelity for the primary objective $J$ was recently examined in \cite{hocker_robustness1_2014}, defined as a convolution of the Hessian $\mathcal{H}_{\beta}(t,t') = \delta^2 J/ \delta \beta(t) \delta \beta(t')$ of $J$ and a noise correlation function $R=\langle \delta \beta(t) \delta \beta(t') \rangle_{\beta}$ characterizing a stochastic parameter in the Hamiltonian $\beta(t)$ ($\beta$ is either $ \epsilon_i$ or $\omega_i)$ subject to noise as $\beta(t) \rightarrow \beta(t) + \delta \beta(t)$. The robustness measure $K_{\beta}$ is the expected change in $J$ due to noise in $\beta$,
\begin{align}
\label{eq:K_a}
K_{\beta} &= \frac{1}{2} \langle \mathcal{H}_{\beta}, R \rangle \nonumber \\
&=\frac{1}{2} \int_0^T \int_0^T \mathcal{H}_{\beta}(t,t') R(t,t') dt dt'.
\end{align}
$K_{\beta}$ values here are negative at high $J$ fidelity, and robustness is enhanced when $K_{\beta}$ is \emph{maximized} towards zero value.

While the degree of robustness depends upon the form of $R$ \cite{hocker_robustness1_2014}, which is often not well known \emph{a priori}, the link to the landscape principle guarantees that the near invariance behavior of $K_{\beta}$ should not significantly depend on the choice of $R$. For demonstration purposes we restrict ourselves to Wide Sense Stationary (WSS) processes whose correlation function decays in time delay $\tau=t-t'$, as these well represent common noise forms found in quantum systems, such as $1/f$, $1/f^2$ and white noise \cite{Martinis_PRL_2007, Koch_pra_2007,Petta_prb_2007,Petta_prb_2007, wineland_nist_1998}. A decaying exponential noise correlation function $R(t,t') = A^{2}\exp \left(-|t-t'|/ \alpha \right)$ is used, with a correlation time of $\alpha=1$ utilized to represent mid-frequency noise ($\alpha = 2$ in Fig. \ref{fig:mcfront}, $\diamond$ curve) \cite{hocker_robustness1_2014}, and the noise strength is arbitrarily chosen as  $A^{2}=10^{-4}$.  For the four-level system, we assume similar but uncorrelated noise statistics from each spin such that robustness is quantified as a sum of contributions from each field or energy level as  $K_\beta = K_{\beta_1} + K_{\beta_2}$. 

Calculations were performed that seek to minimize the error $E_J$ in the primary objective and maximize the robustness measure $K_{\beta}$, where $J$ is chosen as one of the objectives in eqs. (\ref{eq:pif})-(\ref{eq:J}). Evolutionary algorithms  form the base of many multi-objective optimization codes, and here a multi-objective variant of the covariance matrix adaptation method (MO-CMA) was implemented with a population size of 100 members attempting to locate the invariant front (\emph{i.e.}, the highest flat value of $K_{\beta}$) for each type of control objective  \cite{mocma, JPhysA.44.2011.095302}. Such calculations typically required 1000 generations for two-level objectives, and $10^4-10^5$ generations for four-level objectives. The MO-CMA controls were amplitudes $a_k$ and phases $\phi_k$ of a Fourier field structure:

\begin{align}
\label{eq:fields_morerandom}
\epsilon_i(t) &= \sum_k a_k \sin(\omega_k t + \phi_k).
\end{align}

The frequencies were set at  $\omega_k \in [0,10\pi]$ (2-level system) and $\omega_k \in [0, 20\pi]$ (4-level system), and $\Delta \omega_k=\pi$. The initial amplitudes were randomly chosen from either the low fluence regime ($a_k \in [0,0.05]$) or a potentially high fluence regime ($a_k \in [0,50]$), with fluence defined as $K_{f_i} = \int_0^T \epsilon_i^2(t)dt$. As an alternative to the multi-objective search algorithm,  MC solutions were generated from an ensemble of D-MORPH optimization trajectories, performed using random initial fields defined by eq. (\ref{eq:fields_morerandom}). In the D-MORPH optimizations, field values $\epsilon(t)$ at the designated time points were used as controls, in contrast to MO-CMA where amplitudes and phases are used as controls. Parametrized controls are a practical constraint when using MO-CMA, and may impact the maximum $K_{\beta}$ value. Each MC run results in a flat, nearly invariant value for $K_{\beta}$ (Fig. \ref{fig:mcfront}), and the highest such line is taken as the optimal value of $K_{\beta}$ drawn from a large set of MC runs.

\begin{figure}[htb]
\subfloat{\includegraphics[width=0.95\textwidth, height=0.95\textheight, keepaspectratio]{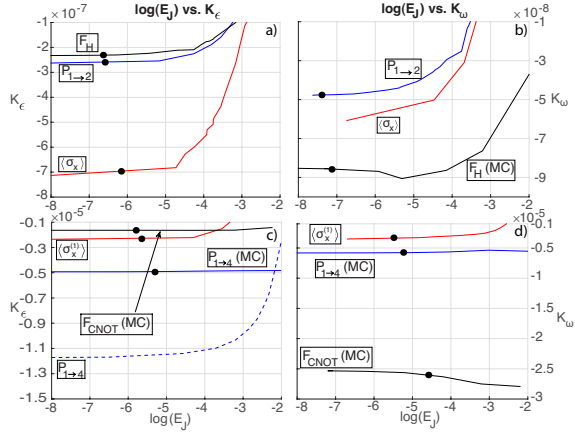}}
\caption{\label{fig:fronts} Color online. MO-CMA and Monte-Carlo (MC) fronts for dual robustness and fidelity objectives. a), c) field noise and b), d) detuning noise of two- and four-level objectives, respectively. Threshold points $E_J = -K_{\beta}$ shown as dots for converged results, beyond which noise effects dictate that the fidelity $J$ cannot be enhanced without first re-engineering the system to improve its noise characteristics.  Dotted line in c) shows MO-CMA performance for comparison with the MC outcome.}
\end{figure}
Figure \ref{fig:fronts} shows the best outcomes between MO-CMA and MC fronts for optimization of $E_J$ and $K_{\beta} $. While a degree of competition is initially present at higher $E_J$, convergence toward invariance is displayed at lower $E_j$ between the two costs for any objective. In common multi-objective optimization problems, MC is typically not an effective algorithm, and it is quite surprising that MC solutions outperformed MO-CMA in several cases, needing only modest numbers of MC solutions  (Fig. \ref{fig:mcfront} inset). This behavior indicates that the top of the quantum control landscape is often replete with robust solutions readily located by moderate number of MC runs.  In a case where MC performed quite well (Fig. \ref{fig:fronts} c), the MO-CMA front is also shown as a dotted line for comparison.  For MC solutions in the invariant regions, low fluence solutions were common for field noise, and high fluence solutions common for detuning noise (not shown). The MO-CMA code located fields of high fluence in both cases.

As mentioned earlier, the implications for the appearance of an invariance plateau suggest that beyond a threshold point $E_J =- K_{\beta}$ depicted by the dots in Fig. \ref{fig:fronts}, there is no benefit in attempting to locate controls that parametrize higher fidelity values of $J$, as the effects of noise would overcome those efforts. The  location of this threshold depends upon the objective being optimized, as well as the strength and nature of the noise. Attempts to go beyond this threshold require re-engineering of the system to have a more favorable noise correlation function such that the threshold can be lowered to allow for a higher fidelity primary objective value for $J$, as demonstrated in Fig. \ref{fig:mcfront}. In a more general context, any primary objective can only be improved beyond its plateaued, invariant secondary threshold value through some suitable form of system engineering.

\begin{figure}[htb]
\subfloat{\includegraphics[width=0.80\textwidth, height=0.80\textheight, keepaspectratio]{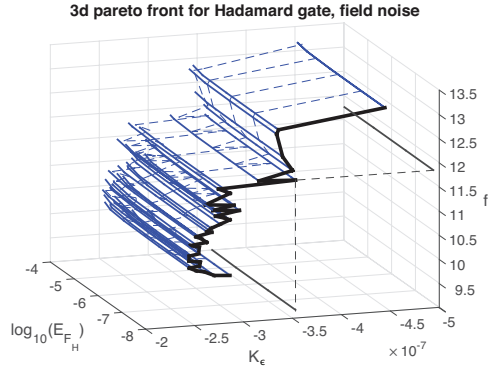}}
\caption{\label{fig:3dfront} Color online. 3D Pareto surface identified using MC trajectories (solid blue lines) for $E_{F_H}$,$K_{\epsilon}$, and $f$ costs. Each MC trajectory shows invariance of a second objective (\emph{i.e.}, projected black lines), but displays complex 2D behavior between the two ancillary objectives (bold line). }
\end{figure}

The invariance behavior between the primary and secondary objective still leaves open likely complex, competitive behavior among the secondary objectives. A MC sampling of such behavior (\emph{i.e.}, the bold front) is shown in Figure  \ref{fig:3dfront} for $E_{F_H}$, $K_{\epsilon}$, and fluence $K_f$ costs.  The invariant behavior will exist for any number of secondary objectives, illustrated here for  $E_{F_H}$ vs. $ K_{\epsilon}$ or $f$ (shown as projected black lines) at all values of the secondary objectives at high fidelity of the primary objective $E_{F_H}$. Interestingly, fields that are optimal for $F_H$ and $K_{\epsilon}$ are not optimal for $F_H$ and $f$, indicated by the fold of the 3D front at low $f$. The general features of the 3D front are only approximately identified in Fig. \ref{fig:3dfront} through modest MC sampling, and including knowledge of the landscape principle within the sophistication of evolutionary codes could possibly improve search efficiency for many secondary objectives.

To summarize, when multi-objective optimizations focus in the regime of high fidelity for a particular primary objective, then the landscape principle assures flat behavior for all secondary objectives that only depend on the control field.  In regards to optimizing robustness alongside fidelity, observing invariance behavior reveals the attractive absence of a major obstacle for advancing quantum technologies, when it is sufficient to operate at $E_J > -K_{\beta}$. If one needs to go beyond the latter point in fidelity, an emphasis must be placed on appropriate further system engineering. The same conclusion would apply more generally to encroachment of any invariant secondary objective where the fidelity passes a practical threshold of undesirable competition with a secondary objective.  This paper dealt with systems undergoing unitary dynamics, and the same principle would apply in cases where at least the significant portion of the local environment is also under full control. Including $K_{\omega}$ in this work is an effective stand-in for that situation. Finally, the invariance phenomenon explained here is not restricted to quantum dynamics, as the same landscape principle is also known to be operative for closed, classical Hamiltonian systems \cite{carlee_2015}. Thus, the principle established in this paper should have wider applicability in multi-objective optimization where a primary objective is demanded to have high, reachable fidelity, and can be exploited in design simulations, as well as directly in the laboratory in an iterative fashion.

\begin{acknowledgements}
The authors would like to thank Ofer Shir and Robert Kosut for their insightful conversations. D.H. acknowledges support from NSF fellowship (DGE 1148900), and H.R. acknowledges support from NSF (CHE-1058644) for landscape principles and ARO-MURI (W911NF-11-1-2068) for quantum information science.
\end{acknowledgements}

\bibliographystyle{apsrev4-1}
\bibliography{hocker}

\end{document}